# Development of high-sensitivity chip calorimeters for cellular metabolic heat sensing

Jihye Kim[a†], Sumin Seo[a†], Jonghyun Kim[a], Sungmin Nam[a], and Wonhee Lee[a, b*]

[a]Graduate School of Nanoscience and Technology, Korea Advanced Institute of Science and Technology (KAIST), Daejeon 34141, Republic of Korea

[b]Department of Physics, KAIST, Daejeon 34141, Republic of Korea

*E-mail: whlee153@kaist.ac.kr

†Authors equally contributed

**Abstract**

Cellular metabolic rate is a good indicator of the physiological state of cells and its changes, which can be measured by total heat flux accompanying metabolism. Chip calorimeters can provide label-free and high throughput measurements of cellular metabolic rate, however, lack of high power resolution and microfluidic sample handling capability has been preventing their wide applications. We report high-resolution chip calorimeters integrated with thin-film parylene microfluidics, which can reliably measure metabolic heat from mammalian cells with controlled stimuli. The molding and bonding technique allowed fast and reliable parylene microfluidic channel fabrications and highly sensitive vanadium oxide thermistor enabled temperature resolution as small as ~ 15 μK, which led to a three-orders-of-magnitude improvement in volume specific power resolutions. Measurements of metabolic heat were successfully demonstrated with adherent and nonadherent cells. We expect the chip calorimeter will provide a universal platform for fundamental cell-biology studies and biomedical applications including cell-based assay for drug discovery.

# Introduction

The metabolic rate of cells can be used for an important parameter that can allow us to gather information on the internal process of the cells[1]. For instance, the basal metabolic rate (BMR) of cells varies depending on the type of the cell and the physiological state of the cell such as malignant transformation. The changes of metabolic rate as responses to external stimuli such as drugs allow us to understand how the stimuli work on cells. Cells produce heat during diverse metabolic processes and the heat production rate of cells is closely related to the metabolic rate of the cells. Therefore, the quantitative measurements of the metabolic heat of cells can provide methods for fundamental studies of cell biology and can be utilized for biomedical applications such as cell-based assay for drug efficacy and toxicity.

Calorimeters have many unique advantages for biosensing applications [2–4]. First of all, calorimetry is a label-free sensing from natural states; not only no fluorescent or radioactive labels are needed but also no additional probe reactants are required. Therefore, there are no risks for cells to be affected by labeling or probe reactants. In addition, heat sensing is insensitive to sample's physical, chemical and optical properties. Moreover, thermal transducers can be completely passivated so that calorimeter is free of biofouling, which causes serious limitations for many other biochemical sensors. Despite these advantages, calorimeters are not widely used for cellular metabolic heat sensing. Commercialized benchtop calorimeters are very sensitive because it has been optimized for measuring the heat of chemical reactions for a long time. However, these calorimeters are very expensive and not easy to be modified for the purpose of metabolic heat sensing. On the other hand, chip calorimeters have advantages in measuring cellular metabolic rate in real-time due to its small sample volume (~nL-µL) and small time constant (~ sec). Microfluidics combined with chip calorimeters can provide many advantages; high throughput, parallel measurements are feasible and diverse microfluidic techniques are available for cell culture and assays within the microfluidic device. In addition, mass-production can allow cheap and disposable device.

The biomedical applications of chip calorimeters reported to date were mostly the measurements of the heat of reactions using biomolecules[5–9]. There were several attempts for the measurement of cellular metabolic heat using chip calorimeters[10–13]. However, the performance of chip calorimeters still requires improvements in several areas. According to the category of chip calorimeters we suggested [2], each type of chip calorimeters has distinct pros and cons. Closed-chamber chip calorimeters have relatively low resolutions while open-chamber chip calorimeters with simple fluidic manipulation, such as pipetting or inkjet, have limitations in the methods for fluid and cell handling. Thin-film parylene microfluidic chip calorimeters made a breakthrough in resolution by applying vacuum-insulation to thin parylene channels [5]. Thermal conductance and device heat capacity (parasitic heat capacity) were dramatically reduce without sacrifice of microfluidic capabilities. However, the previous parylene microfluidic design had a rather small reaction chamber volume for cellular metabolic rate measurement. Unlike biochemical reactions where the heat powers can be easily increased with concentration, metabolic powers from a given number of cells do not change a lot. Therefore, the volume specific power resolution, that is a power resolution divided by measurement chamber volume, becomes an important number to consider. The power resolution of the parylene chip calorimeter was as high as a few nW, however, its sample volume was only a few nL [5]; the resulting volume specific power resolution was in the order of 1 W/L, which should be increased about three orders of magnitude for relatively easy measurements for cellular metabolic rates [14]. The conventional parylene channel fabrication technique with sacrificial photoresist has critical difficulty in making large chamber with long channels because it takes extremely long time for removal of photoresist in such structure. In addition, the larger reaction chamber leads to lager device thermal conductance and lower power resolution.

Here, we report parylene microfluidics chip calorimeters with a large chamber volume and highly-sensitive thermometer for sensitive measurements of metabolic heat from mammalian cells. A novel parylene bonding method using nano-adhesive layers via initiated chemical vapor deposition (iCVD) facilitates fabrication of parylene channel without sacrificial photoresist[15]. In addition, the parylene channels can resist to higher pressure. High-sensitivity thermometry was realized by vanadium oxide ($VO_x$) thermistor which has a high temperature coefficient of ~ -3.8 %/K. We show the feasibility of the chip calorimeter as a novel cell-based assay platform by demonstrating measurements of cellular metabolic heat from both adherent and nonadherent mammalian cell lines.

## Results and discussion

**Device structure and components**

The parylene microfluidic chip calorimeter is composed of three main components: thermometers for monitoring the temperature change at the calorimeter chamber, structures for thermal insulation, and a microfluidic system for handling cells and fluid flows (Figure 1).

As our chip calorimeter is operated in a heat-conduction mode, the resolution of the temperature sensor ($T_{res}$) and the thermal conductance ($G$) between the measurement chamber and the substrate directly affect the resolution ($P_{res}$) of the calorimeter ($P_{res} = T_{res}/G$). A highly sensitive, vanadium pentoxide ($V_2O_5$) thermistor was integrated with the parylene microfluidics as a thermometer [16]. We configured the thermistor in a Wheatstone bridge circuitry with two thermistors for the measurement chamber and the reference chamber [Figure 1(b)]. The heater for thermal conductance calibration was located directly under the measurement chamber. The thermistors were surrounding the measurement chamber to avoid the thermistor degraded by water (Figure 1(b) and (d)). A metallic temperature distributor layer was deposited (~ 70 nm thick silver) and patterned to cover both measurement chamber and thermistor region. The temperature distributor minimizes the temperature gradient within the measurement chamber and reduces the temperature difference between the measurement chamber and the thermistor.

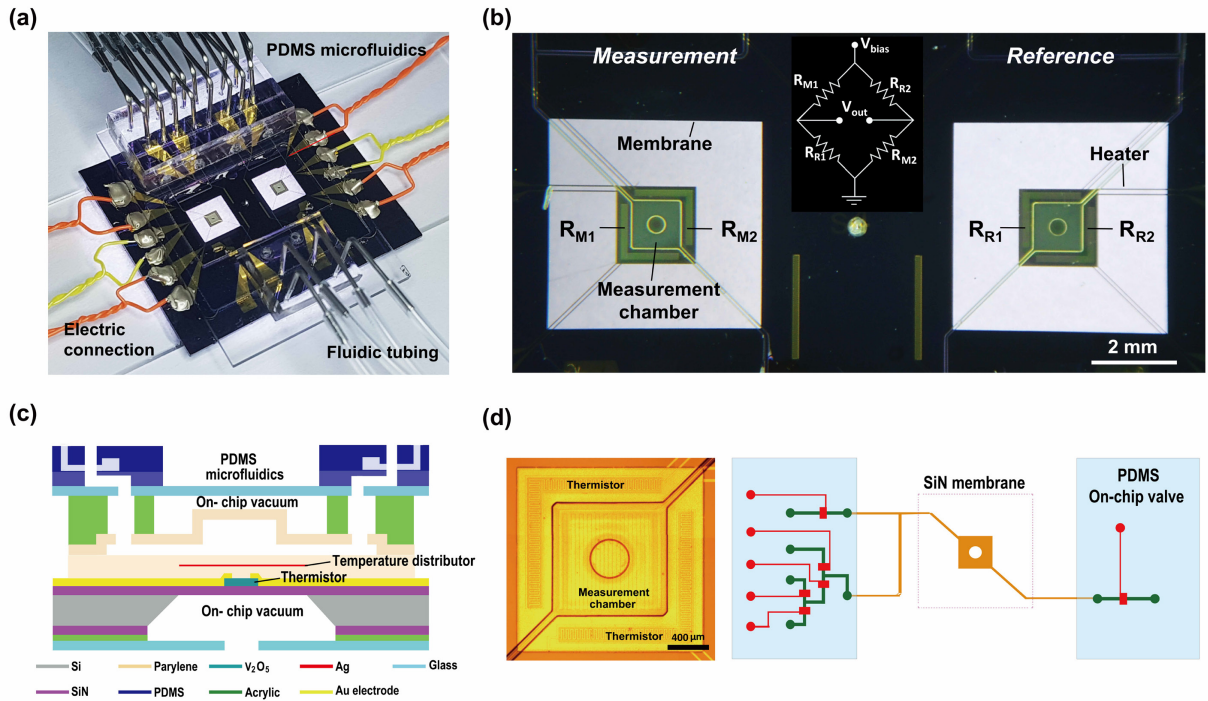

**Figure 1. Microfluidic chip calorimeter device. (a)** Final chip calorimeter device with tubing and electric connections. **(b)** Cross-sectional view. **(c)** Measurement and reference chambers are suspended on SiN membrane. Four thermistors are connected in bride circuit to measure the temperature change. **(d)** PDMS On-chip valves are connected on the parylene channels for cell/drug handling.

The microfluidic system was composed of two distinct parts (: a thin-film parylene microfluidics for enhanced thermal properties of the calorimeter and a two-layer PDMS microfluidics for the fluid manipulation with on-chip valves. The parylene microfluidics made up the calorimeter body including the measurement chamber, the connecting channel and the membrane structure surrounded by the on-chip vacuum chamber. To minimize the thermal conductance, the measurement chamber was built with a thin-film parylene microfluidics and suspended within an on-chip vacuum chamber. The thin-film parylene allows minimal thermal conduction and the vacuum insulation removes conduction and convection through the air that would surround the measurement chamber in a normal condition [5].

As it is difficult to incorporate valves in the parylene microfluidics, the parylene channel was designed as simple as possible and the PDMS microfluidics parts with on-chip valves were connected to the parylene microfluidics to enable the control of loading/removal of cells and flows of fluids including media, buffer, and drugs. The measurement chamber volume was 65 nL (1200 μm × 1200 μm × 50 μm) excluding center circular post (radius of 200 μm) for parylene channel's mechanical stability. There is some possibility that cells or cell debris remain in the channel after cell injection, especially at the parylene-PDMS connection part where some dead volume is unavoidable due to parylene-PDMS channel alignment issue. To avoid such issue, one of the inlets was used as designated cell inlet while the other one was used for fluid inlet. The valves on the fluid inlet side were closed and cells were flowed in the measurement chamber. Then the cell inlet could be washed with buffer backflow from fluid inlet and the cell inlet was remain closed during the measurements. By separating the cell inlet and fluid inlet, we can also avoid issues of potential cross-contaminations. The fluid inlet of the parylene was further branched in the PDMS part to use multiple fluids. The current design has 3 fluid inlets for media and two different drug solutions. The on-chip valve at the outlet was closed during the cell inlet washing and kept open otherwise.

**Device fabrication**

The chip calorimeter device fabrication steps are shown in Figure 2. A SiN wafer was used for the substrate. The SiN layer was 200 nm thick and was patterned on the backside for KOH etching step. Then the $V_2O_5$ thermistor was fabricated by optical lithography and annealing (Figure 2(a), step 1) [16]. First, a 200-nm-thick $VO_x$ was sputter-deposited (Ar and $O_2$ flow rate of 20 sccm : 5 sccm) and patterned by lift-off. Then, the VO$_x$ was annealed in air environment at (400 °C for 3 h) to achieve the V$_2$O$_5$ phase. Electrodes were fabricated by sputter-depositing Cr (7 nm) and Au (70 nm). Parylene coating is often used as a passivation layer for sensors and electronic devices owing to its great barrier property for gas and moisture. However, the parylene layer of a-few-micrometer thickness was not sufficient for completely blocking the water vapor transmission. The thermistor was located outside the measurement chamber to evade the water damage issue. Silver temperature distributor (~ 70 nm thickness) was deposited on the area including the measurement chamber and surrounding thermistors (Figure 2(b), step 2). As a protective layer for electrodes, 5 μm thick parylene layer was deposited by a commercial parylene coater (SCS Labcoater 2, Specialty coating system, USA).

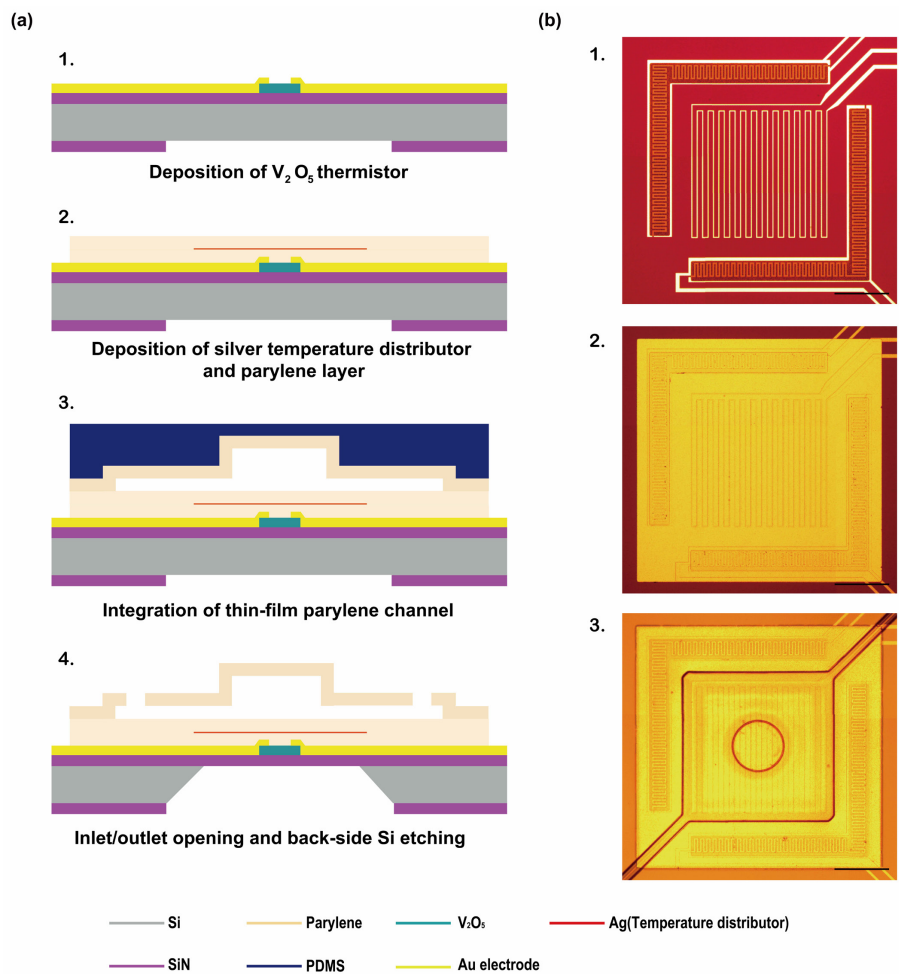

**Figure 2. Fabrication procedure. Schematic (a) and corresponding device picture (b) 1. V$_2$O$_5$ thermistor is constructed by photolithography and sputtering followed by annealing. 2. Parylene passivation layer is deposited with parylene coater. Silver layer is deposited to distribute temperature uniformly. 3. Parylene channel is molded and bonded with nano-adhesive layer deposited via iCVD. 4. Membrane is released by bulk silicon etching and in/outlet is constructed for fluid injection (scale bar, 400 μm).**

Parylene microfluidic channel was integrated on top of the thermometers by molding and bonding technique (Figure 2(a), step 3)[15]. Parylene was deposited on a PDMS channel mold that was replicated from an SU-8 mold. The parylene layers on the substrate and the PDMS mold were then bonded by nanoadhesive layer. A poly-glycidyl methacrylate (PGMA) layer of 200 nm thickness was deposited on both of the substrates by initiated chemical vapor deposition (iCVD). One of the two PGMA coated parylene substrate was reacted with ethylene diamine (Sigma Aldrich) in a petri dish at 75 °C for 5 min. Then the two parylene layers were bonded at 80 °C vacuum oven for 8 hours. The PDMS mold substrate could be easily removed after the bonding was completed. Compared to the conventional fabrication method involving photoresist sacrificial layer, bonding and molding allows faster fabrication of microfluidic designs with complex and long channel geometry. Especially the design of the calorimeter has a large chamber at the center that is connected with a long channel. Removal of sacrificial photoresist in such geometry is extremely impractical as the dissolution of photoresist can take several weeks. Moreover, the parylene channel can often delaminate with pressure from fluid flow and the surrounding vacuum. The high bonding strength of the iCVD bonding improves the resistance to channel delamination. In addition, parylene's mechanical properties, such as the intrinsic stress, can change with high temperature fabrication processes as a result of thermal annealing of parylene. The molding and bonding process prevents such issue because the process is performed below the glass transition temperature of parylene (~90 °C for parylene C).

The measurement chamber area was suspended into a membrane by bulk Si etching with 30 % KOH solution at 83 °C (Figure 2(a), step 4). On-chip vacuum chambers and inlet/outlets were constructed by bonding acrylic plates (1 mm thick) and slide glasses (1 mm thick) on the top and bottom side (Figure 1(b)). The acrylic plates were laser-cut to a shape of vacuum chamber and bonded to the chip calorimeter device with UV adhesive (NOA 71, Norland products). For the top side acrylic plate, inlet/outlet holes were also made with laser cutting. The inlet/outlet holes on the top glass slide were made with a diamond drill. A vacuum pumping hole was made between the measurement chamber and the reference chamber to connect the on-chip vacuum spaces at the top and the bottom of the chip. PDMS on-chip valve was composed of a 30-μm-thick flow layer and a 5-mm-thick valve layer. The channel height of the flow layer was 20 μm. The PDMS on-chip valve was bonded to the slide glass via air plasma bonding.

**Device characterization**

Temperature resolution of the thermistor was determined by temperature coefficient ratio (TCR) and electrical noise. The TCR of the $V_2O_5$ thermistor was measured with the probe station in the temperature range of 30-80 °C (Figure 3(a)). The measured value of the TCR was - 3.8 %/K. The resistance of the thermistor was 35 kΩ at 37 °C, and the resistivity was 15 Ω·m. It was also confirmed that there was no Metal-Insulation-Temperature (MIT) or hysteresis at the measured temperature range. The electric noise of the thermistor and the measurement system was measured with varying bias voltage (Figure 3(b)). The RMS voltage noise was determined by measuring the output voltage of the bridge circuit with a lock-in amplifier (SR830, Stanford Research Systems). The voltage noise was measured at 0.001-1 Hz bandwidth while increasing the bias voltage from 0.1 V to 1.3 V. The voltage noises at low bias voltages (0.1-0.2 V) were close to Johnson noise level of 24 nV/√Hz. The noise RMS voltage increased linearly with the bias voltage, which is known property of 1/f noise with bias voltage increase. Noise equivalent temperature difference (NETD) can be calculated as NETD = $1/2 \times V_{noise} / (V_{bias} \cdot TCR)$ for the bridge circuitry used for the measurements (Figure 1(b)). At the low bias voltage where Johnson noise dominates, NETD was relatively large. With the increasing bias voltage, 1/f noise portion increases and eventually will dominate over other noises. The NETD of the thermistor bridge decreased and converged to the minimum value of ~ 5 μK, which can be explained as the 1/f noise behavior because the NETD is proportional to $1/V_{bias}$ while 1/f noise is proportional to $V_{bias}$. The resolution or detection limit of the temperature sensor ($T_{res}$) is ~ 15 μK (signal to noise ratio = 3:1) when $V_{bias} > 0.5$ V.

The device thermal conductance was measured by applying an electric heat pulse to the heater integrated at the measurement chamber (Figure 3(c)). Electric heating of varying power ($P$) was applied to the heater using a function generator (B2962A, Keysight). The output voltages were measured at the plateau and the temperature changes ($\Delta T$) were calculated. The thermal conductance ($G$) was calculated by the slope of the linear fit ($1/G=\Delta T/P$) and was 28 μW/K. The inset graph shows the output voltage responding to a 1 μW heating pulse. This thermal conductance value was similar to the value of the device developed previously (16 μW/K), while the chamber volume of the current device was increased almost 20 times (from 3.5 nL to 66 nL) [5]. The thermal relaxation time constant of the device was calculated by fitting a exponential curve and was ~ 8.5 s.

The power resolution of the device was calculated with the thermal conductance and temperature resolution. We achieved a high power resolution of 420 pW by improving the temperature resolution. As the application target of the chip calorimeter is the metabolic heat of cells, the heat source is roughly proportional to the number of cells and the sample volume becomes an important factor to be considered. The average basal metabolic rate of mammalian cells is known to be ~ 30 pW [17]. The power resolution of 420 pW would be a reasonable number for measurement of metabolic power of tens of cells.

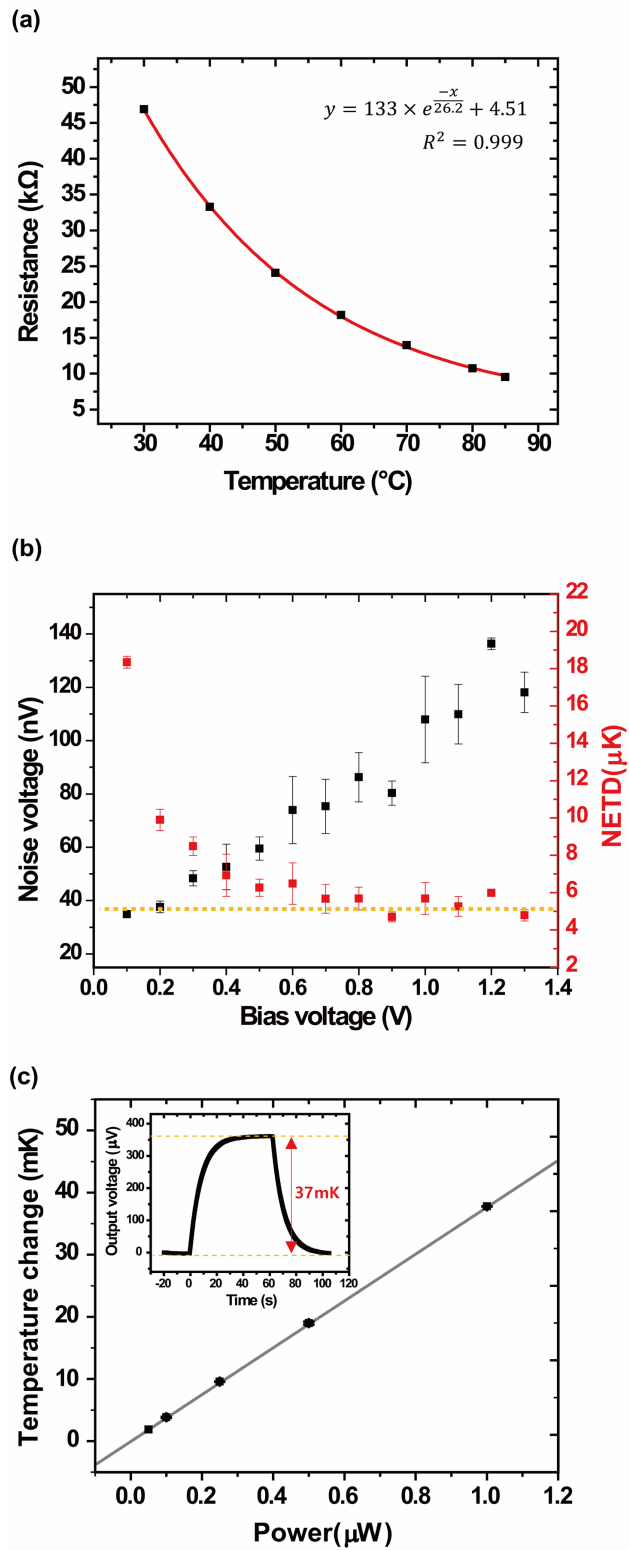

**Figure 3.** Chip calorimeter performance. (a) TCR of thermistor: -3.8 %/K. (b) Noise voltage (RMS) of the thermistor bridge circuit and NETD with varying bias voltage. (c) Thermal conductance of the chip calorimeter (1/G= ΔT/P). Inset : Output voltage changes with applied power of 1 μW.

Typical cell cultures using culture dishes have cell density in the range ~$10^6$-$10^7$ cells/mL. Microfluidic cell culture platforms can typically have ~ one order of magnitude larger cell density with better control of mass transport. We would need at least 10 nL for 100 cells to have ~ $10^7$ cells/mL density. The larger the chamber volume, we can have the more number of cells that can be supported reliably during the measurements. Here we significantly increased the chamber volume and improved the power resolution. The resulting volume specific powerresolution can be defined as $P_{res}/V$ to quantify the performance of the chip calorimeter for applications in measurements of cellular metabolic heat. The volume specific power resolution of the chip calorimeter was 6.4 pW/nL, which is improved by 3 orders of magnitude from the previous device [5]. The table 1 summarizes the comparisons and the improvements.

**Table 1. Chip calorimeter performance**

|  | Previous Device [5] | Current Device |
|---|---|---|
| Thermal conductance ($G$) | 16 µW/K | 28 µW/K |
| Time constant ($\tau$) | 1.3 s | 8.5 s |
| Chamber volume ($V$) | 3.5 nL | 65.7 nL |
| Temp. resolution ($T_{res}$) | 260 µK | 15 µK |
| Power resolution ($P_{res}$) | 4.2 nW | 420 pW |
| Volume specific power resolution ($P_{res}/V$) | 1.2 nW/nL | 6.4 pW/nL |

**Measurements of cellular metabolic heat**

Measurements of cellular metabolic rate were demonstrated with adherent and nonadherent cell types; HeLa cell was used for the adherent cell type and Jurkat T-cell was used for the nonadherent cell type. The interior surfaces of the parylene microfluidic channel after the bonding have active epoxy groups on the top surface and active amine groups on the bottom side. The amine groups can be used for immobilizing various proteins on the surface. For the measurements using adherent cells, fibronectin (Sigma Aldrich) was coated on the chamber surface before the cell loading. First, EDA was flown in the channel to react with remaining epoxy groups. Then microfluidic channel was incubated at 37 °C for 5 minutes for the reaction and washed thoroughly with PBS. Then, the fibronectin solution (1mg/ml in PBS) was flowed in the channel and incubated at 37 °C for 30 minutes. For nonadherent cells, bovine serum albumin (Sigma Aldrich) was coated on the chamber surface to avoid non-specific binding of unwanted proteins and cells on the measurement chamber surface. Similar to the fibronectin immobilization procedure, the chip calorimeter was incubated at 37 °C for 30 minutes after the injection of bovine serum albumin (1 wt % in PBS buffer). After the surface coating on the measurement chamber with appropriate biomolecules, the chip calorimeter was loaded in the measurement box, where the chip calorimeter was connected with vacuum pumping, fluidic tubing, and electric connections. The measurement box was put in an environmental box with a microscope. The environmental box was maintained at 37 °C.

Figure 4(a) shows the measurement of metabolic heat changes during lysis of norepinephrine-stimulated HeLa cells. Cells were prepared at a concentration of $3.0 \times 10^7$ cells/ml and injected into the chamber. With the measurement chamber volume of 65 nL, the injected number of cells within the measurement chamber was expected to be ~2000 on the average. Here, the number of prepared cells was large to make the metabolic heat signal as large as possible. They were incubated for 1.5 hours for cell adhesion to the bottom surface (Figure 4(b) inset). Media with 10 μM norepinephrine was injected to stimulate the cells. Then, Lysis buffer (10 % w/v Sodium dodecyl sulfate in distilled water) was injected to induce cell death. The inlet pressure on the flow channel was 4.4 psi, which produces the flow rate of ~ 3 μL/min in our device design. Lysis buffer was injected for 2 minutes to inject enough amount for parylene microfluidic channel volume. The entire parylene channel and measurement chamber have ~ 0.6 μL total volume and inlet/outlet holes at acrylic plate and glass slide have ~1.5 μL total volume. There is a period that the measured power seems to drop rapidly right after the buffer injection (~ 2-4 min in Figure 4(a)). This is an artifact due to the temperature drop by injection of external fluids. The temperature of the measurement chamber is higher than everywhere else because of the cellular metabolic heat power and self-heating of the thermistor. A sudden influx of external fluids would lead to temperature drop at the measurement chamber. Aside from the temperature drop by fluidic injection, the metabolic power was observed to stabiliz at a lower value after ~ 8 min, which represents the difference due to the absence of metabolic heat from cells. Cell death after the lysis buffer injection was confirmed by the microscope. Exposure to such high concentration of lysis buffer causes fast cell death. The cell lysis was tested in the same condition as the calorimetric measurement was done and the cell death was observed within a minute using a cell viability assay using 20 μM calcein AM. The measurement of the metabolic rate showed the difference of ~12 μW before and after the cell death. This value corresponds to ~ 5.8 nW per cell. Although the cells were activated with norepinephrine, such value is unexpectedly high considering average BMR of single mammalian cells is ~ 30 pW [17]. Cellular metabolic power measured to date were done with tissue samples and there must be a difference in metabolic rate between tissue samples and culture cells. Moreover, it is well known that cancer cells have significantly higher metabolic rate than normal cells. However, the difference of two orders of magnitude was surprising and further comprehensive investigations are required to verify the source of the high metabolic rate.

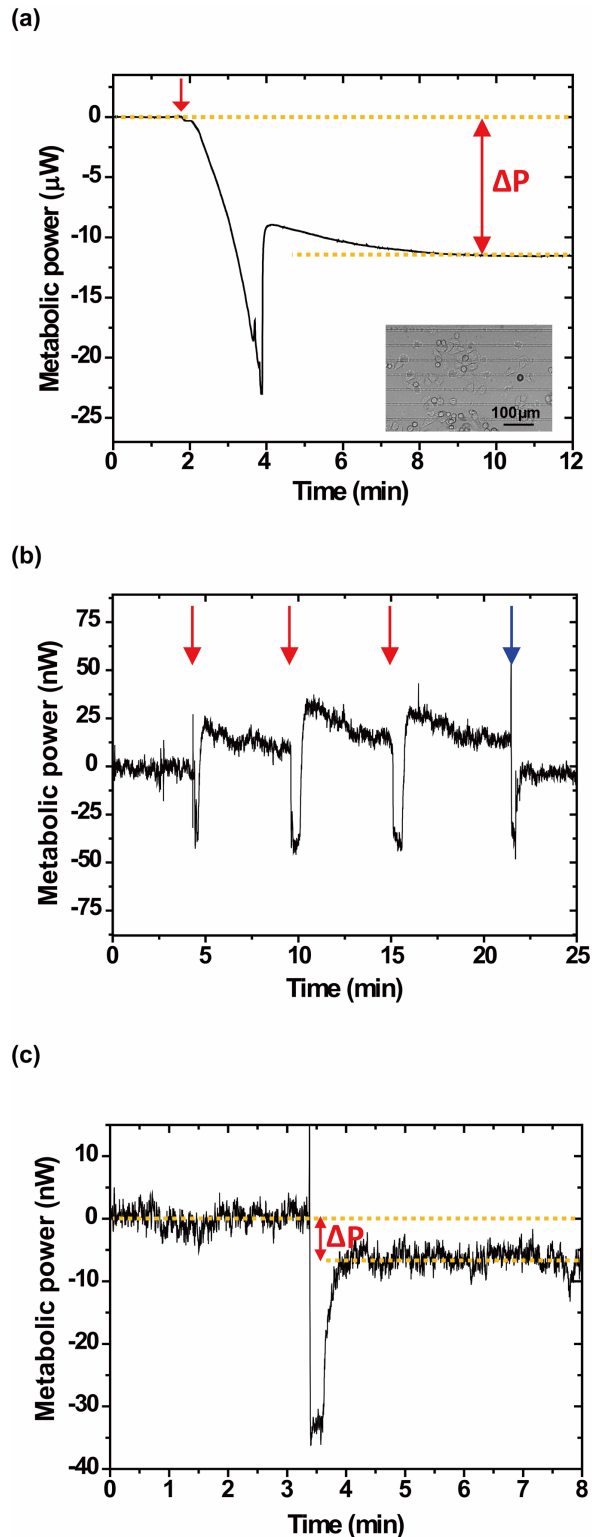

**Figure 4.** Measurements of cellular metabolic heat power. (a) Metabolic power change over cell lysis. Hela cells were exposed to lysis buffer at ~ 2 min (red arrow). Cells are attached and spread inside the measurement chamber before the measurement (inset). (b) Metabolic power change with injection of 10 μM norepinephrine (red arrow) and cell culture media (blue arrow). Number of cells was ~ 200. (c) Metabolic power of T cells. The cells in the measurement chamber was washed out at ~3.3 min. Number of cells were ~ 200.

Real-time response to norepinephrine stimulation was tested with ~ 200 cells (Figure 4(b)). Cells were prepared in the same way as before but with less concentration. The cells were exposed to the injections of norepinephrine (10 μM in cell culture media) at 5, 10, 15 min (red arrow) and the measurement chamber was flushed with culture media (Dulbecco's Modified Eagle's Media, Welgene) at 22 min (blue arrow). Norepinephrine and culture media were injected for 30 seconds. There were fluid injection signals observed again. After the transient fluid injection signal faded away, the increased metabolic heat power could be observed. The metabolic power increased by norepinephrine stimulation and then slowly decreased. We believe the metabolic energy was mainly limited by the oxygen in the chamber and decreased faster than under normal media. With another injection of norepinephrine, the level of the stimulated metabolic rate recovered and repeated the decreasing pattern again. When the chamber was flushed with culture media, the metabolic power returned to the value before the injection of norepinephrine. These results show that the metabolic rate responses to the norepinephrine immediately after the injection and removal. Our chip calorimeter can detect the metabolic rate changes responding to biochemical stimuli almost in real time. There is the period that the signal cannot be measured, which corresponds to the period of the transient signal after the injection. The transient signal last for ~ 40 s, which includes the period of injection/mixing and thermal relaxation time.

It is relatively simple to treat the adherent cells with drugs while keeping the cells in the chamber and measuring the metabolic heat. On the other hand, is not a trivial problem to keep non-adherent cells in the measurement chamber during the injection of the drug or culture media. We prepared suspended cells mixed with the desired concentration of drug before introducing the cells in the measurement chamber. The metabolic heat power was measured while the cells suspended in culture media mixed with the drug were injected all together into the chamber and washed out by culture media. Numbers of the cells in the measurement chamber were counted with a microscope.

Metabolic heat of Jurkat T-cell was measured with norepinephrine stimulation (Figure 4(c)). Metabolic heat was monitored while the suspended T cells stimulated with norepinephrine were injected into the chamber and flushed by culture media (Roswell Park Memorial Institute 1640 media, Welgene). Figure 4(c) shows the metabolic heat change when the cell suspension mixed with 10 μM norepinephrine was washed out from the measurement chamber. The metabolic heat power when the cells were in the chamber was set as zero for convenience. The metabolic heat power dropped to - 6.9 nW after washing out. The number of cells in the chamber was 228 and 42 remained after flushing the chamber. Therefore, we can deduce that metabolic heat per cell was 37 pW. Similar measurements with 20 μM norepinephrine resulted in metabolic heat power change per cell of ~ 260 pW, showing cellular metabolic heat power was increased with the higher concentration of norepinephrine.

## Conclusion

Microfluidic chip calorimeters integrated with thin-film parylene microfluidics were developed for the measurements of cellular metabolic heat. New molding and bonding technique allowed reliable and fast parylene microfluidic channel fabrication, which led to a significant increase in measurement chamber volume. Highly sensitive vanadium oxide thermistor and low noise measurement circuitry allowed high temperature resolution of ~ 15 μK. The chip calorimeter provides power resolution of ~ 420 pW that was sufficient for reliable measurement of the BMR of tens of mammalian cells. We successfully demonstrated cellular metabolic heat measurement with adherent and nonadherent cells. Despite the long history of calorimetry, the metabolic heat sensing from cells are still at its infancy. We anticipate the chip calorimeter will provide a general platform for fundamental studies in cell biology and for many biomedical applications including cell-based assay for drug discovery.